\documentclass[prb,twocolumn,aps,showpacs,fixfloats,minimal]{revtex4}
\usepackage{graphicx}
\usepackage{bm}
\usepackage{amsmath,amssymb}
\usepackage{subfigure}
\usepackage{float}
\usepackage{latexsym}
\usepackage{epstopdf}
\usepackage{color}
\usepackage{enumerate}
\usepackage{pdfpages}

\begin{document}
\newcommand{\s}{\scriptscriptstyle}
\newcommand{\uu}{\uparrow \uparrow}
\newcommand{\ud}{\uparrow \downarrow}
\newcommand{\du}{\downarrow \uparrow}
\newcommand{\dd}{\downarrow \downarrow}
\newcommand{\ket}[1] { \left|{#1}\right> }
\newcommand{\bra}[1] { \left<{#1}\right| }
\newcommand{\bracket}[2] {\left< \left. {#1} \right| {#2} \right>}
\newcommand{\vc}[1] {\ensuremath {\bm {#1}}}
\newcommand{\tr}{\text{Tr}}
\newcommand{\Trans}{\ensuremath \Upsilon}
\newcommand{\Refl}{\ensuremath \mathcal{R}}

\title{Spectral narrowing and spin echo for localized carriers with  heavy-tailed L{\'e}vy distribution of hopping times}

\author{Z. Yue$^{1}$, V. V. Mkhitaryan$^{2}$,  and M. E. Raikh$^{1}$ }

\affiliation{$^{1}$Department of Physics and
Astronomy, University of Utah, Salt Lake City, UT 84112, USA \\
$^{2}$Ames Laboratory, Iowa State University, Ames, Iowa 50011, USA}

\begin{abstract}
We study analytically the free induction decay and the spin echo
decay originating from the localized carriers
%for the localized carriers
moving between the sites which
host random magnetic fields.
%hosting random magnetic fields in the regime
%when
Due to disorder in the site positions and energies, the on-site residence times,
$\tau$, are widely spread according to the L{\'e}vy distribution.
The power-law tail $\propto \tau^{-1-\alpha}$ in the  distribution
of $\tau$ does not affect the conventional spectral narrowing for $\alpha >2$,
but leads to a dramatic acceleration of the free induction decay in the domain $2>\alpha >1$.
The next abrupt acceleration of the decay takes place as $\alpha$ becomes smaller than $1$.
In the latter domain the decay does not follow a simple-exponent law.
%behavior.
To capture the behavior of the average spin in this domain, we solve the
evolution equation for the average spin using the approach different from
the conventional approach based on the Laplace transform. Unlike the free
induction decay, the tail in the distribution of the residence times leads
to the slow decay of the spin echo. The echo is dominated by realizations
of the carrier motion  for which the number of sites,
visited by the carrier, is minimal.
\end{abstract}
\maketitle
\section{Introduction}
A concept of the spectral narrowing of the magnetic resonance lineshape
was quantified more than sixty years ago  in a seminal paper Ref. \onlinecite{Anderson}.

\begin{figure}
\includegraphics[width=70mm]{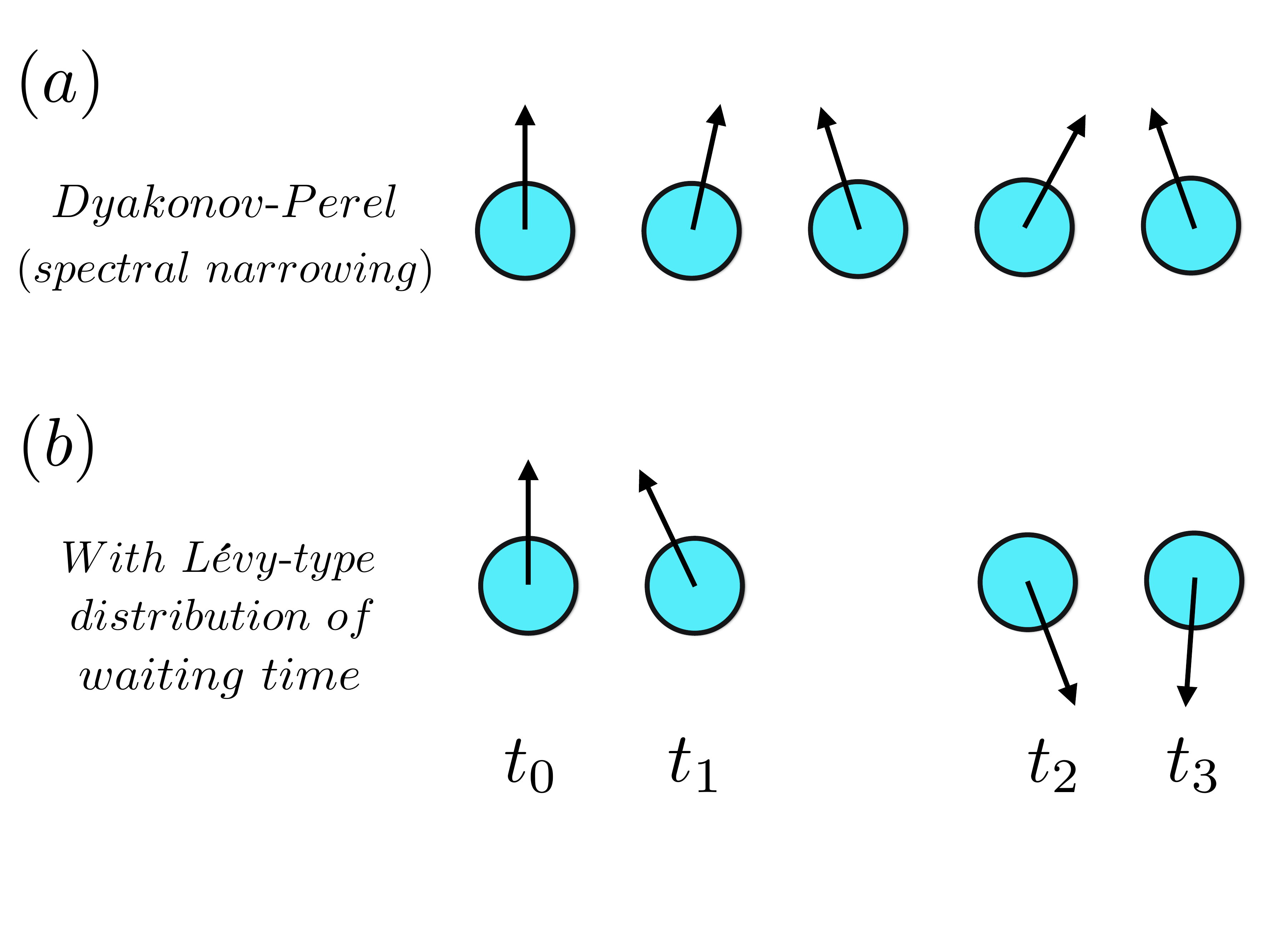}
\caption{(Color online)
The contrast between (a) Dyakonov-Perel spin relaxation with a single correlation time and (b) spin relaxation with broad distribution of the waiting times is illustrated schematically. Allowance for anomalously long
waiting times accelerates the relaxation.}
\label{figure1}
\end{figure}

In application to free induction decay (FID), this concept can be recapped as follows.
In the presence of the time-dependent random magnetic field, $b(t)$, the  decay of the  FID signal is  determined by the average $\big\langle\exp\Big[i \int\limits_0^t dt' b(t')\Big]\big\rangle$. The character of the decay
 depends on the relation between the typical magnitude, $b_{\s 0}$, of $b(t)$  and the correlation time, $\tau$. For long
correlation time $b_{\s 0}\tau \gg 1$, the decay is gaussian, $\propto \exp(-b_{\s 0}^2t^2)$, reflecting the gaussian distribution
of the magnitudes of $b(t)$. In the opposite limit, $b_{\s 0}\tau \ll 1$,  the integrand rapidly changes sign, which is
the origin of the spectral narrowing. If the time intervals between the subsequent sign changes are, $\delta t_1$,
$\delta t_2$, $\delta t_3$, and so on, then the average,  $\big\langle \exp\Big[i\int\limits_0^t dt' b(t')\Big]\big\rangle$, over the field realizations
can be
%estimated
rewritten as
$\exp\Big[-b_{\s 0}^2\sum\limits_{\s i= 1}^n (\delta t_i)^2\Big]$. On the other hand, the number of the sign changes, $n$, is determined by
the condition $\sum\limits_{\s i= 1}^n \delta t_i=t$. This leads to a simple exponential behavior, $\exp(-t/\tau_{\s s})$, of the FID signal, where
\begin{equation}
\label{FID}
\frac{1}{\tau_{\s s}}=b_{\s 0}^2\frac{\overline{ (\delta t)^2}}{\overline{  \delta t }}.
\end{equation}
If the random field is characterized by a {\em single} correlation time, $\tau_{\s 0}$, then the intervals $\delta t_i$
obey the Poisson distribution
\begin{equation}
\label{Poisson}
p_{\s 0}(\delta t,\tau_{\s 0})=\frac{1}{\tau_{\s 0}}\exp\big[-\delta t/\tau_{\s 0}\big].
\end{equation}
Averaging with this distribution in Eq. (\ref{FID}) yields a well-known result,
$\tau_{\s s}=1/2b_{\s 0}^2\tau_{\s 0}$, for the decay rate.
In the field of semiconductors this result is also known
as the Dyakonov-Perel spin relaxation time\cite{DP1971}.

A nontrivial situation emerges when the correlation times, $\tau$, are broadly distributed.
Then  Eq.~(\ref{FID}) takes the form
\begin{equation}
\label{FID1}
\frac{1}{\tau_{\s s}}=2b_{\s 0}^2\frac{\langle \tau^2\rangle}{\langle \tau\rangle   },
\end{equation}
where the averaging
is performed over the distribution, $F(\tau)$, of the correlation times.  Such a situation is generic, e.g.,
for the dispersive transport in disordered semiconductors\cite{DT1,DT2,DT3,DT4,Harmon1,Harmon2,Baranovskii}.
Broad distribution of the $\tau$-values stems from the spread in the activation energies. Another example
is a system with hopping transport, where the broad distribution of $\tau$ is the result of the
spread in the hopping distances. In both cases $F(\tau)$ has a power-law
tail: $F(\tau)\propto \tau^{-1-\alpha}$. Such a distribution, also known as the
L{\'e}vy distribution, is normalizable for positive $\alpha$.
However, for $\alpha <2$ the average $\langle \tau^2\rangle$ diverges.
Formally, this implies that  $\tau_{\s s}$ turns to zero.
On the physical level, this means that,  on certain occasions,
the spin spends enough time in some given  field to exercise a full rotation,
see Fig. \ref{figure1}.  Although the portion of these occasions is small
$\sim \left(b_{\s 0}\tau_{\s 0}\right)^{\alpha}$, they change
the average spin dynamics dramatically. Theoretical study of this
dynamics is the subject of the present paper. We find that, for $\alpha <2$,
the FID retains the form of a simple exponent, but the rate, $\tau_{\s s}^{-1}$,
shortens and becomes a strong function of the tail parameter, $\alpha$.
Our results can be summarized as
\begin{equation}
\label{cases}
\frac{1}{\tau_{\s s}}=\begin{cases}
 D_1(\alpha)~b_{\s 0}^2\tau_{\s 0},~~~~~~~~~~ \alpha>2 \\
 D_2(\alpha)~b_{\s 0}^{\alpha}\tau_{\s 0}^{\alpha-1},~~~~~ 2> \alpha >1\\
 D_3(\alpha)~b_{\s 0},~~~~~~~~~~~~1> \alpha>0,
 \end{cases}
\end{equation}
where $D_1(\alpha)$, $D_2(\alpha)$, and  $D_3(\alpha)$ are the dimensionless
functions of the tail parameter.
Change of the behavior of $\tau_s$ at $\alpha=2$ is due to
the formal divergence of $\langle \tau^2\rangle$, while the
change at $\alpha =1$ is due to the formal divergence of
$\langle \tau\rangle$, which enters into the denominator of Eq. (\ref{FID1}).
We also find that the crossovers at $\alpha=2$ and $\alpha =1$ take
place within narrow intervals:
$|\alpha -2|\sim 1/|\ln b_{\s 0}\tau_{\s 0}|$
and $|\alpha -1|\sim 1/|\ln b_{\s 0}\tau_{\s 0}|$.

Another phenomenon which is strongly affected in the presence of multiple
waiting time is spin echo\cite{KlauderAnderson}. The effect of the tail
in $F(\tau)$ on the echo is opposite to the effect
of $F(\tau)$ on the spin relaxation rate.
The echo decays {\em slower} due to this tail. The average echo
signal is determined by the realizations with longest waiting times.

The paper is organized as follows. In Sect. II we derive
a closed equation for the FID averaged over the realizations of
random fields. Asymptotic (in parameter $b_{\s 0}\tau_{\s 0} \ll 1$)
solution of this equation is found in Sect. III, where we derive
the result Eq. (\ref{cases}) and also find the crossover behaviors
near $\alpha=2$ and $\alpha=1$.  In Sect. IV we analyze the decay of the
average echo signal.
Concluding remarks are presented in Sect. V.

%Then the question arises: what happens to the spin dynamics, $\langle S_z(t)\rangle$, if the {\em typical} $\tau$ satisfies the condition $b_{\s 0}\tau \ll 1$, but the average $\tau$ diverges? Such a divergence takes place when the distribution, $F(\tau)$, of the $\tau$-values
%has a slow-decaying tail: $F(\tau)\propto \tau^{-1-\alpha}$ with $\alpha <1$.
%A physical consequence of this tail is that, on certain occasions, the spin spends enough time to exercise the full rotation, see Fig. \ref{figure1}.  Although the portion of these occasions is small $\sim \left(b_{\s 0}\tau_{\s 0}\right)^{\alpha}$, they change
%the average spin dynamics completely.
%
%{DT1,DT2,DT3,DT4,Harmon1,Harmon2,Baranovskii}

\section{Basic equation for average FID with multiple relaxation times}
As it was mentioned in the Introduction, the physical picture
which we have in mind is the carrier motion between the
sites either by hopping or by trapping-detrapping process\cite{DT1,DT2,DT3,DT4,Harmon1,Harmon2,Baranovskii}.
The time-dependent magnetic field of a hyperfine origin\cite{hyperfine}, acting on the carrier spin,
represents a sequence of steps
${\bm b}(t)=\sum_i{\bm b_i}\left[\Theta(t_{i+1}-t)-\Theta(t-t_i)\right]$,
where $\Theta(x)$ is the step-function and the step durations,
$(t_{i+1}-t_{i})$, are distributed according to the Poisson distribution
Eq. (\ref{Poisson}), in which $\tau_{\s 0}$ is distributed according to  $F(\tau)$.
Since the sites are separated in space, random fields, ${\bm b}_i$,
at different sites are {\em completely uncorrelated}.
We also assume that the motion is three-dimensional,
so that the effect of occasional returns to the
same site\cite{Czech,Robert1,Mkhitaryan}  are negligible.

Suppose that at time moment $t=0$ a carrier occupies the
site $i=0$, and its spin is directed along the $z$-axis.
After time $t$ the carrier can either remain on the site $i=1$
or hop on the neighboring site $i=1$. The probability
to stay is $p_{\s 0}(t,\tau^{\s (0)})$, defined by Eq. (\ref{Poisson}),
where $\tau^{\s (0)}$ is a waiting time for the hop
$0\rightarrow 1$. We assume that the external magnetic field
is directed along $x$-axis, so that only
the $x$-components of the fields $b^{\s (0)}$
and $b^{\s (1)}$ on the sites $i=0$ and $i=1$ are important.
If the carrier stays on $i=0$, then the $z$-projection of
its spin after time $t$ is equal to $\cos b^{\s (0)}t$.
If the carrier hops after time $t_{\s 1}<t$, then this projection
is equal to $\cos\left[b^{\s (0)}t_{\s 1}+b^{\s (1)}(t-t_{\s 1})\right]$.
Taking into account that the moments $t_{\s 1}$ are random,
the value of $S_z(t)$ can be presented as a sum
\begin{align}
\label{0}
S_z(t)=&p_{\s 0}(t,\tau^{\s (0)})\cos b^{\s (0)}t \\ \nonumber
&+\int\limits_0^t dt_{\s 1}\frac{d(1-p_{\s 0}(t_{\s 1},\tau^{\s (0)}))}{dt_{\s 1}}\cos\left[b^{\s (0)}t_{\s 1}+b^{\s (1)}(t-t_{\s 1})\right].
\end{align}
The derivative in the integrand is the probability density of the hop.
If there is a site $i=2$ on which the carrier can
hop from $i=1$, the expression Eq. (\ref{0}) gets modified.
It acquires a third term describing the possibility of the
hop $1\rightarrow 2$, with corresponding waiting time $\tau^{\s (1)}$.
If this hop takes place, $S_z$ acquires the value
$\cos\left[b^{\s (0)}t_{\s 1}+b^{\s (1)}t_{\s 2}+b^{\s (2)}(t-t_{\s 1}-t_{\s 2})\right]$,
where $t_{\s 2}$ is a random residence time on the site $i=1$ and $b^{\s (2)}$
is the random field on the site $i=2$.

For infinite number of possible hops Eq. (\ref{0}) transforms
into an infinite series. Averaging each term over the gaussian
distribution of magnetic fields and realizations of waiting
times generates a series for average spin projection,
$\overline{S}_z(t)$. It can be
%checked
verified that $\overline{S}_z(t)$ satisfies the equation
%Vagharsh, your equation for $\overline{S}_z(t)$ reads
\begin{equation}
\label{1}
\overline{S}_z(t)=\Big\langle e^{-t/\tau}\Big\rangle e^{-b_{\s 0}^2t^2}+\int\limits_0^t dt_1A(t_1) \overline{S}_z(t-t_1),
\end{equation}
where the function $A(t)$ is defined as
\begin{equation}
\label{2}
A(t)=\Big\langle\frac{1}{\tau} e^{-t/\tau}\Big\rangle e^{-b_{\s 0}^2t^2},
\end{equation}
and $\langle ....\rangle$ stands for averaging over the broadly distributed waiting
times. The above closed equation describes the averaged spin relaxation. In the next Section
we solve it in different domains of the tail parameter $\alpha$.

\section{Solution of Equation for ${\overline S}_z(t)$}
For concreteness, we choose the distribution function of the waiting times in the form
\begin{equation}
\label{F}
F(\tau)=\frac{C_\alpha \tau_{\s 0}^\alpha}{(\tau_{\s 0}^2+\tau^2)^\frac{1+\alpha}{2}}.
\end{equation}
For $\tau \gg \tau_{\s 0}$, we have $F(\tau)\propto \tau^{-1-\alpha}$, and the coefficient $C_{\alpha}$ which insures the
normalization is given by
\begin{equation}
\label{C}
C_\alpha=\frac{2\Gamma(\frac{1+\alpha}{2})}{\pi^{1/2}\Gamma(\frac{\alpha}{2})}.
\end{equation}
As we demonstrate in the Appendix, for the multiple trapping model\cite{Baranovskii} the form Eq. (\ref{F}) describes accurately not only the tail but the
entire body of the distribution.
While the {\em typical} time, $\tau_{\s 0}$, is short, $\tau_{\s 0}\ll b_{\s 0}^{-1}$,
%\begin{equation}
%\label{3}
%b\tau_{\s 0}\ll 1,
%\end{equation}
the distribution
%of $\tau$
has a long tail $F(\tau)\propto \tau^{-1-\alpha}$.

We start the analysis of the basic equation  Eq. (\ref{1}) by noticing that at times $t\gg \tau_{\s 0}$ the first term  is small. Indeed, averaging with the help of Eq. (\ref{F}), yields
\begin{equation}
\label{4}
\Big\langle e^{-t/\tau}\Big\rangle \sim \Big(\frac{\tau_{\s 0}}{t}\Big)^{\alpha}.
\end{equation}
A crucial step of the analysis is making use of the fact that
spin relaxation takes place over a large number of hops.
This allows one to expand
$\overline{S}_z$ in the integrand
\begin{equation}
\label{5}
\overline{S}_z(t-t_1)\approx \overline{S}_z(t)-t_1\frac{d\overline{S}_z}{dt}(t).
\end{equation}
Upon  this expansion,  Eq. (\ref{1}) can be easily solved yielding
\begin{equation}
\label{6}
\overline{S}_z(t)=\exp\Big[-\int\limits_0^tdt_1\Phi(t_1)\Big],
\end{equation}
where the function $\Phi(t)$ is defined as
\begin{equation}
\label{7}
\Phi(t)=\frac{1-\int\limits_0^tdt_2A(t_2)}{\int\limits_0^t dt_2t_2A(t_2)}.
\end{equation}
%We specify the distribution function of the form
%\begin{equation}
%\label{F}
%F(\tau)=\frac{C_\alpha \tau_{\s 0}^\alpha}{(\tau_{\s 0}^2+\tau^2)^\frac{1+\alpha}{2}}.
%\end{equation}
%For $\tau \gg \tau_{\s 0}$, we have $F(\tau)\propto \tau^{-1-\alpha}$, and the coefficient $C_{\alpha}$ which insures the
%normalization is given by
%\begin{equation}
%\label{C}
%C_\alpha=\frac{2\Gamma(\frac{1+\alpha}{2})}{\pi^{1/2}\Gamma(\frac{\alpha}{2})}.
%\end{equation}
For $\alpha >1$ the characteristic time, $\tau_s$, for the spin dynamics is much longer than $b_{\s 0}^{-1}$ (we will check this assumption later). On the other hand, even for wide distribution of the waiting times, the function $A(t)$ falls off dramatically for  $t\gtrsim b_{\s 0}^{-1}$. This allows one to extend the
upper limits in the integrals in Eq. (\ref{7}) to $\infty$.
Then Eq. (\ref{6}) reduces to a simple exponential decay
\begin{equation}
\label{8}
\overline{S}_z(t)=\exp\Big(-\frac{t}{\tau_{\s s}}\Big),
\end{equation}
with the decay rate given by
\begin{equation}
\label{9}
\frac{1}{\tau_{\s s}}=\Phi(\infty)=\frac{\int\limits_0^{\infty}dt\Big\langle\frac{1}{\tau} e^{-t/\tau}\Big\rangle \Big(1-e^{-b_{\s 0}^2t^2}\Big)}{\int\limits_0^{\infty}dt t\Big\langle\frac{1}{\tau} e^{-t/\tau}\Big\rangle e^{-b_{\s 0}^2t^2}}.
\end{equation}
If the averages in the numerator and denominator decayed rapidly at $t\gtrsim \tau_{\s 0}$,
we would be allowed, by virtue of the condition $b_{\s 0}\tau_{\s 0} \ll 1$, to replace $\exp\big(-b_{\s 0}^2t^2\big)$ by $1$ in the denominator and expand
$\Big[1-\exp\big(-b_{\s 0}^2t^2\big)\Big]$ in the numerator.  In this way,
we would retrieve the standard expression Eq. (\ref{FID1})
for the Dyakonov-Perel relaxation time.
%\begin{equation}
%\label{10}
%\frac{1}{\tau_{\s s}}=b^2\frac{\Big\langle \tau^2\Big\rangle}{\Big\langle \tau\Big\rangle}
%\end{equation}
Calculating $\langle \tau^2\rangle$ and $\langle \tau\rangle$
with the help of the distribution Eq. (\ref{F}) we find

\begin{equation}
\label{11}
\frac{1}{\tau_{\s s}}= \frac{\pi^{1/2}(\alpha-1)\Gamma(\frac{\alpha-2}{2})}{2\Gamma(\frac{\alpha+1}{2})} b_{\s 0}^2\tau_{\s 0}
\end{equation}
This expression is valid if $\langle \tau^2\rangle$ is finite, which corresponds to $\alpha>2$.
The prefactor in this expression specifies the function $D_1(\alpha)$ in Eq. (\ref{cases}). The function
$D_1(\alpha)$ falls off monotonically with $\alpha$. At $\alpha \gg 1$ it behaves as $\Big(\frac{2\pi}{\alpha}\Big)^{1/2}$.

In the domain $1<\alpha <2$ the value of $\langle \tau^2\rangle$ diverges while  $\langle \tau\rangle$ remains finite. The latter still allows one to set $b_{\s 0}=0$ in the denominator of Eq. (\ref{9}), but the numerator cannot be expanded anymore.
The explicit expression for the numerator in Eq. (\ref{9}) reads

 \begin{align}
\label{12}
\int\limits_0^{\infty}dt\Big\langle\frac{1}{\tau} e^{-t/\tau}\Big\rangle\Big(1-e^{-b_{\s 0}^2t^2}\Big)=\\ \nonumber
\int\limits_0^{\infty}d\tau\int\limits_0^{\infty}dt\frac{C_\alpha \tau_{\s 0}^\alpha}{(\tau_{\s 0}^2+\tau^2)^\frac{1+\alpha}{2}\tau}e^{-t/\tau}\Big(1-e^{-b_{\s 0}^2t^2}\Big)
\end{align}
Upon introducing the new variables $z=t/\tau$ and $w=b_{\s 0}z\tau$, the integral in the right-hand side takes the form
\begin{equation}
\label{13}
 C_\alpha (b_{\s 0}\tau_{\s 0})^\alpha\int\limits_0^{\infty}dz ~z^\alpha e^{-z}  \int\limits_0^{\infty}dw
 \frac{1-e^{-w^2}}{[(b_{\s 0}\tau_{\s 0}z)^2+w^2]^{\frac{\alpha+1}{2}}}.
\end{equation}
Since the characteristic $z$ in Eq. (\ref{13}) is $\sim 1$, we can neglect $(b_{\s 0}\tau_{\s 0}z)^2$ in the denominator, after which the double integral factorizes, yielding
\begin{equation}
\label{14}
 C_\alpha (b_{\s 0}\tau_{\s 0})^\alpha\Gamma(\alpha)\Gamma\Big(\frac{2-\alpha}{2}\Big).
\end{equation}
Then the corresponding expression for the relaxation time in the domain $1<\alpha<2$
acquires the form
\begin{equation}
\label{15}
\frac{1}{\tau_{\s s}}=(\alpha-1)\Gamma(\alpha)\Gamma\Big(\frac{2-\alpha}{2}\Big)  \frac{(b_{\s 0}\tau_{\s 0})^\alpha}{\tau_{\s 0}},
\end{equation}
where the prefactor specifies the function $D_2(\alpha)$ in Eq.~(\ref{cases}).
\subsection{Vicinity of $\alpha=2$}

 We see that at the demarkation value $\alpha=2$ both functions $D_1(\alpha)$ and $D_2(\alpha)$ diverge, so that the expressions Eq. (\ref{11}) and Eq. (\ref{15}) yield $\tau_{\s s}\rightarrow 0$. This suggests that the crossover domain should
 be treated more carefully. Namely, we rewrite the integral over $w$ in Eq. (\ref{13}) as
 \begin{align}
\label{16}
&\int\limits_0^{\infty}dw
 \frac{1-e^{-w^2}}{[(b_{\s 0}\tau_{\s 0}z)^2+w^2]^{\frac{\alpha+1}{2}}}=\frac{\Gamma(\frac{2-\alpha}{2})}{\alpha}
 \\\nonumber &+\int\limits_0^{\infty}dw\Big[\frac{1}{[(b_{\s 0}\tau_{\s 0}z)^2+w^2]^{\frac{\alpha+1}{2}}}
 -\frac{1}{w^{\alpha+1}}\Big]\Big(1-e^{-w^2}\Big).
\end{align}
The integral in the right-hand side converges at small $w \sim b_{\s 0}\tau_{\s 0}z$, which allows one to expand $(1-e^{-w^2})$. Upon introducing the variable $v=(\frac{w}{b_{\s 0}\tau_{\s 0}z})^2$, this integral takes the form
\begin{equation}
\label{17}
\frac{(b_{\s 0}\tau_{\s 0}z)^{2-\alpha}}{2}\int\limits_0^{\infty}dv\Big[\frac{v^{1/2}}{(1+v)^{\frac{\alpha+1}{2}}}
 -\frac{1}{v^{\frac{\alpha}{2}}}\Big].
\end{equation}
Now we rewrite the integral $\int\limits_0^\infty \frac{dv}{v^{\frac{\alpha}{2}}}$ as a sum
of integrals from $0$ to $1$ and from $1$ to $\infty$. The integral from $1$ to $\infty$
is then combined with the first integral in Eq. (\ref{17})  in which domain of integration
is shifted by $1$.  This yields

\begin{equation}
\label{18}
\frac{(b_{\s 0}\tau_{\s 0}z)^{2-\alpha}}{2}\Big[-\int\limits_0^{1}dv\frac{1}{v^{\frac{\alpha}{2}}}
+\int\limits_1^{\infty} dq \frac{(q+1)^{1/2}-q^{1/2}}{q^{\frac{\alpha+1}{2}}}\Big].
\end{equation}
Note that the second integral in the square brackets remains finite at $\alpha=2$, while the first
integral diverges. Keeping only the diverging part and combining it with 
$\frac{1}{\alpha}\Gamma\Big(\frac{2-\alpha}{2}\Big)\approx \frac{1}{2-\alpha}$,
we establish the behavior of the spin relaxation rate Eq. (\ref{15}) near $\alpha=2$ 
\begin{equation}
\label{19}
\frac{1}{\tau_{\s s}}= 2b_{\s 0}^2\tau_{\s 0}|\ln b_{\s 0}\tau_{\s 0}|\Upsilon\Big((\alpha-2)|\ln b_{\s 0}\tau_{\s 0}|\Big).
\end{equation}
where the crossover function $ \Upsilon(z)$ is defined as
\begin{equation}
\label{upsilon}
\Upsilon(z)=\frac{1-e^{-z}}{z}.
\end{equation}
Thus the expressions Eq. (\ref{11}) and Eq. (\ref{15}) are valid outside the interval
$|\alpha-2|\sim |\ln b_{\s 0}\tau_{\s 0}|^{-1}$, which is parametrically narrow.

\subsection{Vicinity of $\alpha=1$}
We see that Eq. (\ref{15}) yields $\tau_{\s s}\rightarrow \infty$ as $\alpha$ approaches
$1$ from the above. This is the result of the divergence of $\langle \tau \rangle$
in this limit. To regularize the behavior of  Eq.~(\ref{15}), we need to calculate
the denominator in Eq.~(\ref{9})  more accurately. We start from the explicit  form of this denominator
\begin{equation}
 \label{20}
\int\limits_0^\infty dt tA(t)= \int\limits_0^{\infty}d\tau\int\limits_0^{\infty}dt\frac{C_\alpha \tau_{\s 0}^\alpha}{(\tau_{\s 0}^2+\tau^2)^\frac{1+\alpha}{2}\tau}te^{-t/\tau}e^{-b_{\s 0}^2t^2}
\end{equation}
The same change of variables $z=t/\tau$ and $w=b_{\s 0}z\tau$, allows to cast  the integral in the  the form
\begin{equation}
\label{21}
 C_\alpha \tau_{\s 0}(b_{\s 0}\tau_{\s 0})^{\alpha-1}\int\limits_0^{\infty}dz ~z^{\alpha} e^{-z}  \int\limits_0^{\infty}dw
 \frac{w e^{-w^2}}{[(b_{\s 0}\tau_{\s 0}z)^2+w^2]^{\frac{\alpha+1}{2}}}.
\end{equation}
Formally, the singular behavior of this integral at $\alpha=1$ follows from the fact that
at $\alpha =1$ integration over $w$ yields logarithm if we neglect a small parameter
$(b_{\s 0}\tau_{\s 0})^2$ in the denominator.
To capture this  behavior, we rewrite the integral over $w$ using the
integration by parts

\begin{align}
\label{22}
\int\limits_0^{\infty}dw
 \frac{w e^{-w^2}}{[(b_{\s 0}\tau_{\s 0}z)^2+w^2]^{\frac{\alpha+1}{2}}}=\\ \nonumber
 \frac{1}{\alpha-1}\Big[
  (b_{\s 0}\tau_{\s 0}z)^{1-\alpha}-
 \int\limits_0^{\infty}dw \frac{2we^{-w^2}}{ [(b_{\s 0}\tau_{\s 0}z)^2+w^2]^{\frac{\alpha-1}{2}}}
 \Big].
\end{align}
Now we can safely neglect $(b_{\s 0}\tau_{\s 0})^2$ in the denominator and perform the integration
over $z$, which yields
  \begin{equation}
\label{23}
\frac{C_\alpha \tau_{\s 0}(b_{\s 0}\tau_{\s 0})^{\alpha-1}}{\alpha-1}\Big[\Gamma(2\alpha-1)(b_{\s 0}\tau_{\s 0})^{1-\alpha}-\Gamma(\alpha)\Gamma(\frac{3-\alpha}{2})\Big].
\end{equation}
Finally, the behavior of $\tau_{\s s}$  in the vicinity of $\alpha=1$ can be expressed in terms of the crossover function
$\Upsilon(z)$ defined by Eq. (\ref{upsilon}) 
 \begin{align}
\label{24}
\frac{1}{\tau_{\s s}}&=
\frac{(\alpha-1)\Gamma(\alpha)\Gamma\Big(\frac{2-\alpha}{2}\Big)}
{\Gamma(2\alpha-1)-\Gamma(\alpha)\Gamma(\frac{3-\alpha}{2})(b_{\s 0}\tau_{\s 0})^{\alpha-1}}\frac{ (b_{\s 0}\tau_{\s 0})^{\alpha}}{\tau_{\s 0}} \\ \nonumber
&\approx \frac{(\alpha-1)\pi^{1/2}}
{1-(b_{\s 0}\tau_{\s 0})^{\alpha-1}}\frac{ (b_{\s 0}\tau_{\s 0})^{\alpha}}{\tau_{\s 0}}
=\frac{\pi^{1/2} b_{\s 0}}{|\ln b_{\s 0}\tau_{\s 0}|} \frac{1}{\Upsilon \Big((1-\alpha)|\ln b_{\s 0}\tau_{\s 0}|\Big)}.
\end{align}
Unlike Eq. (\ref{19}), the crossover function appears in the denominator.
Directly at $\alpha=1$ Eq. (\ref{24}) yields
 \begin{equation}
\label{25}
\frac{1}{\tau_{\s s}}= \pi^{1/2}\frac{b_{\s 0}}{|\ln b_{\s 0}\tau_{\s 0}|}.
\end{equation}
The fact that for all $\alpha$ greater or equal $1$ the value  $\tau_{\s s}^{-1}$ is smaller than $b_{\s 0}$  justifies the ansatz:
$\overline{S}_z(t-t_1)=\overline{S}_z(t)-t_1d\overline{S}_z/dt$ and the extension of the upper limit in Eq. (\ref{7}) to $\infty$.

\begin{figure}
\includegraphics[width=85mm]{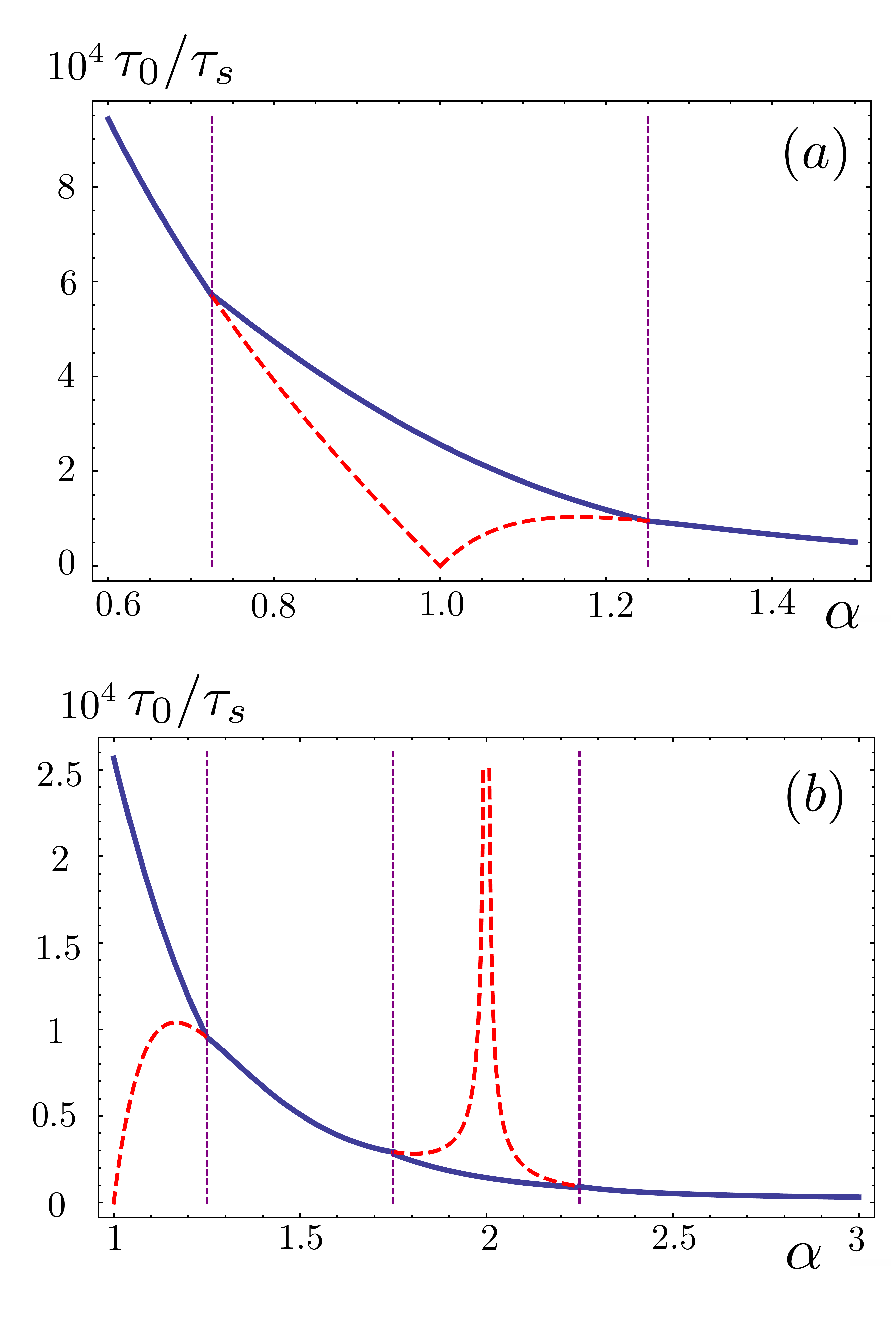}
\caption{(Color online) (a)
Acceleration of the relaxation rate with decreasing the tail-parameter, $\alpha$, is illustrated in the domain $0.6<\alpha<1.5$. Dashed red curves are plotted from Eq. (\ref{D3}) (for $\alpha <1$), and Eq. (\ref{15})  (for $1<\alpha <1.5$). The curves exhibit a dip near $\alpha=1$. They are connected by the crossover expression Eq.  (\ref{24}).
(b) Domain $1<\alpha<3$. Dashed red curves are plotted from Eq. (\ref{15})  (for $1<\alpha <2$),
and from Eq. (\ref{11}) (for $\alpha>2$). The curves exhibit  a divergence near $\alpha=2$. They are connected by the crossover expression Eq. (\ref{19}). All the curves are plotted for $b_{\s 0}\tau_{\s 0}=10^{-3}$.}
 %Three red curves are plotted from Eq. (\ref{D3}) (for $\alpha <1$), Eq. (\ref{15})  (for $1<\alpha <2$),
%and from Eq. (\ref{11}) (for $\alpha>2$). The curves exhibit  a divergence near $\alpha=2$. In this domains the crossover expressions Eq. (\ref{24}) and Eq. (\ref{19}), shown with blue curves, apply. All the curves
%are plotted for $b_{\s 0}\tau_{\s 0}=10^{-3}$.}
\label{figure2}
\end{figure}

\subsection{$\alpha<1$}

For $\alpha<1$ we cannot extend the limits of integration in Eq. (\ref{7}) to infinity.
Instead, we rewrite the expression Eq. (\ref{2}) for $A(t)$ as follows

 \begin{equation}
\label{26}
A(t)=\Big\langle\frac{1}{\tau} e^{-t/\tau}\Big\rangle\Big(e^{-b_{\s 0}^2t^2}-1\Big)
+\Big\langle\frac{1}{\tau} e^{-t/\tau}\Big\rangle.
\end{equation}
The integration over $t$ in the second term can be carried out explicitly.
Then  Eq. (\ref{7}) acquires the form
 \begin{equation}
\label{27}
\Phi(t)=\frac{\Big\langle e^{-t/\tau}\Big\rangle+\int\limits_0^tdt_2\Big\langle\frac{1}{\tau} e^{-t_2/\tau}\Big\rangle\Big(1-e^{-b_{\s 0}^2t_2^2}\Big)}
{\int\limits_0^t dt_2t_2\Big\langle\frac{1}{\tau} e^{-t_2/\tau}\Big\rangle e^{-b_{\s 0}^2t_2^2}}.
\end{equation}
In calculating both averages we can use the power-law tail of  the distribution Eq. (\ref{F})
 \begin{equation}
\label{28}
\Big\langle\frac{1}{\tau} e^{-t/\tau}\Big\rangle\approx C_\alpha\frac{\tau_{\s 0}^{\alpha}}{t^{\alpha+1}}\Gamma(\alpha+1),~~
\Big\langle e^{-t/\tau}\Big\rangle\approx C_\alpha\frac{\tau_{\s 0}^\alpha}{t^\alpha}\Gamma(\alpha).
\end{equation}
This expressions apply for $t\gg \tau_{\s 0}$.
Substituting them into Eq. (\ref{27}) and then Eq. (\ref{28}) into Eq. (\ref{6}),
we arrive at the final result for $\overline{S}_z(t)$
\begin{equation}
\label{29}
\overline{S}_z(t)=\exp\Bigg[-\int\limits_{b_{\s 0}\tau_{\s 0}}^{b_{\s 0}t} dx
\frac{\frac{1}{x^\alpha}+\alpha \int\limits_{0}^{x} dx_1 x_1^{-\alpha-1}(1-e^{-x_1^2})}{\alpha \int\limits_{0}^{x} dx_1 x_1^{-\alpha}e^{-x_1^2}}\Bigg].
\end{equation}
To analyze  Eq. (\ref{29}) in the domain $\tau_{\s 0} <t< b_{\s 0}^{-1}$, we note that
for small $x$ the integrand in the exponent behaves as $\frac{1-\alpha}{\alpha x}$ which
leads to the power-law decay of $\overline{S}_z(t)$

\begin{equation}
\label{30}
\overline{S}_z(t)\Big|_{t\ll b_{\s 0}^{-1}}\approx \exp\Big[-\frac{1-\alpha}{\alpha}\ln(t/\tau_{\s 0})\Big]
=\Big(\frac{\tau_{\s 0}}{t}\Big)^{\frac{1-\alpha}{\alpha}}.
\end{equation}
This behavior should be compared to the first term in Eq. (\ref{1}), which we neglected.
The power-law Eq. (\ref{30}) dominates at $\alpha >0.6$.

In the long-time limit $b_{\s 0}t\gg 1$ the integral in Eq. (\ref{29}) is determined by large $x$.
For large $x$ the integrand  saturates at the value
$\frac{(1-\alpha)\Gamma(\frac{2-\alpha}{2})}{\alpha\Gamma(\frac{3-\alpha}{2})}$, so that the resulting
expression for $\overline{S}_z(t)$ reads

\begin{equation}
\label{31}
\overline{S}_z(t)\Big|_{t\gg b_{\s 0}^{-1}}\approx
\Big(b_{\s 0}\tau_{\s 0}\Big)^{\frac{1-\alpha}{\alpha}}
\exp\Big[-\frac{(1-\alpha)\Gamma(\frac{2-\alpha}{2})}{\alpha\Gamma(\frac{3-\alpha}{2})}
b_{\s 0}t\Big].
\end{equation}
Since we neglected the first term in Eq. (\ref{1}), the result Eq. (\ref{31}) does not capture
 the initial stage of the  decay, which is dominated by this first term.
 The decay in the entire time domain is described by Eq. (\ref{31}) when $\alpha$ is so close to $1$ that the
 prefactor in Eq. (\ref{31}) is not  small. Then Eq. (\ref{31}) captures the spin decay which is a simple exponent with
 \begin{equation}
 \label{D3}
 \frac{1}{\tau_{\s s}}\approx (1-\alpha)\pi^{1/2}b_{\s 0}.
\end{equation}
Comparing to Eq. (\ref{cases}), we conclude that the function $D_3(\alpha)$ has the form $(1-\alpha)\pi^{1/2}$   near $\alpha=1$. Note, that this expression is consistent with Eq. (\ref{24}) for $(\alpha-1)\gg |\ln(b_{\s 0}\tau_{\s 0})|^{-1}$ .
%
%\subsection{Correction from the first term in Eq. {\ref{1}}}
%
%At times $t \gg \tau_{\s 0}$ when the ansatz $\overline{S}_z(t-t_1)=\overline{S}_z(t)-t_1d\overline{S}_z/dt$
%is applicable, the integral equation Eq. (\ref{1}) can be solved analytically with the free term,
%$\langle e^{-t/\tau}\rangle \exp(-b^2t^2)$, taken into account.
%The solution reads
%\begin{align}
%\label{32}
%\overline{S}_z(t)&=\exp\Big[-\int\limits_0^tdt_1\Phi(t_1)\Big]
%\\ \nonumber &
%+\int\limits_0^t dt_1\frac{\Big\langle e^{-t_1/\tau}\Big\rangle e^{-b^2t_1^2}}{\int\limits_0^{t_1} dt_2t_2\Big\langle\frac{1}{\tau} e^{-t_2/\tau}\Big\rangle e^{-b^2t_2^2}} \exp\Big[-{\int\limits_{t_1}^{t}dt_2\Phi(t_2)}\Big].
%\end{align}
%Correction from the free term is described by the second term in the curly brackets.
% The relevant values of $t_1$ in the integrand are $t_1\lesssim b^{-1}$. For $\alpha >1$ this
% restriction allows to neglect the last exponent in the integrand and extend the upper limit
% of the integral in denominator to infinity. The Eq. (\ref{32}) simplifies to
%\begin{align}
% \label{33}
%\overline{S}_z(t)&=\exp\Big[-\int\limits_0^tdt_1\Phi(t_1)\Big]
%\Bigg\{1+\frac{(\alpha-1)\Gamma(\alpha)}{\tau_{\s 0}}\int\limits_0^t dt_1 \Big(\frac{\tau_{\s 0}}{t_1}\Big)^\alpha e^{-b^2t_1^2}\Bigg\}.
%\end{align}
%We see that the correction grows with time and at large times saturates at the value  $(\alpha-1)\Gamma(\alpha)\Gamma(\frac{1-\alpha}{2})/2 (b\tau_{\s 0})^{\alpha-1}$

\begin{figure}
\includegraphics[width=70mm]{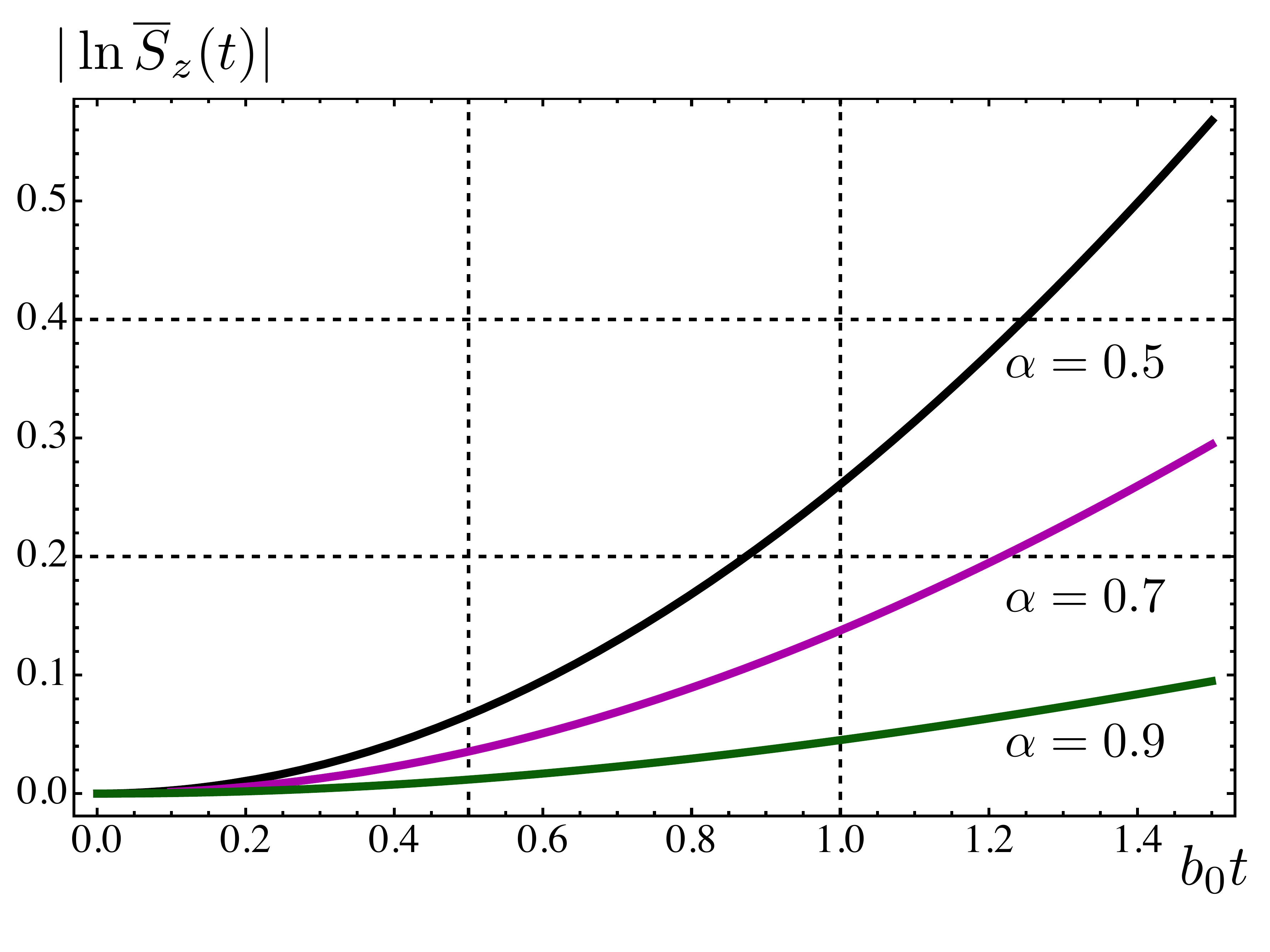}
\caption{(Color online) The time evolution of  $\vert\ln\overline{S}_z(t)\vert$ is plotted from  Eq. (\ref{29}) for
$b_{\s 0}\tau_{\s 0}=10^{-3}$ and the values of the tail-parameter: $\alpha=0.9$ (green), $\alpha=0.7$ (purple), and $\alpha=0.5$ (black).
The smaller is $\alpha$, the later the curves converge to the straight lines.  The convergence is final only for $\alpha=0.9$.
For smaller $\alpha$ the slopes keep  increasing with time beyond
the maximal $t =1.5/b_{\s 0}$ and saturate at the values predicted by Eq. (\ref{31}) only at very large
times. The domain where the first term in Eq. (\ref{1}) dominates the decay corresponds to $b_{\s 0}t\lesssim 10^{-3}$, and
is not represented in the figure.
}
\label{figure3}
\end{figure}

The overall behavior of the relaxation rate with $\alpha$ is illustrated in Fig. \ref{figure2}. In Fig. \ref{figure3}
we plot the time evolution of  $\ln\overline{S}_z(t)$ using the general expression Eq. (\ref{29}). We see that, the smaller
is $\alpha$, the later the curves converge to the straight lines corresponding to a simple  exponential  behavior. The convergence is final only for biggest $\alpha=0.9$. For smaller $\alpha$-values the slopes keep increasing with time beyond
the maximal $t =1.5/b_{\s 0}$ shown in the figure. The slopes saturate at the values predicted by Eq. (\ref{31}) only at very large
times.

\section{Spin echo decay with a L{\'e}vy-type waiting times distribution}
It is well known\cite{KlauderAnderson} that the motion narrowing strongly affects the decay of the spin echo,
which is formally defined as ${\overline S}_E(T)=\big\langle\exp\Big[i \Big(\int\limits_0^{\frac{T}{2}} dt b(t)-\int\limits_{\frac{T}{2}}^T dt b(t)\Big)\Big]\big\rangle$.
If the random magnetic field is characterized by a single correlation time $\tau$, the decay of the spin
echo would follow the FID signal, i.e. ${\overline S}_E(T)={\overline S}(T)$.
The situation is very different for the broad distribution of $\tau$. Indeed, the shortening of the FID time,
$\tau_{\s s}$, for distribution Eq. (\ref{F}) with power-law tail was due to the possibility for a carrier to occasionally sit on a given site, $i$, during the time, $\tau_{\s i}$, much longer than a typical time, $\tau_{\s 0}$.
These sites give anomalously strong contribution, $\big\langle\exp\Big[i \int\limits_{t_{\s i}}^{t_{\s i}+\tau_{\s i}} dt b_{\s i}(t)\Big]\big\rangle$, to the decay. The same physics suggests that, for distribution of $\tau$ with power-law tail, the echo signal will decay {\em slowly} with $T$. This is because
the contributions from anomalously long residence times are eliminated in the echo signal.
To quantify this statement, consider a situation when a carrier populates a certain site, $i$, at time
$T_{\s 1}<T/2$ and leaves it at time $T_{\s 2}>T/2$.  Then the contribution from this site to the unaveraged
echo signal is given by 
\begin{equation}
\label{echo1}
\exp\Big\{i b_{\s i}\Big[\Big(\frac{T}{2}-T_{\s 1}\Big)-\Big(T_{\s 2}-\frac{T}{2}\Big)\Big]\Big\}=
\exp\Big[i b_{\s i}\Big(T-T_{\s 1}-T_{\s 2}\Big)\Big].
\end{equation}
Assume now, that the carrier makes many hops before arrival to the site $i$ and many hops after departure
from the site $i$. Then the probabilities to preserve spin during the time intervals $(0,T_{\s 1})$
and $(T_{\s 2},T)$ are given by $~~~$  $\exp\big[-T{\s 1}/\tau_{\s s}\big]$ and
$\exp\big[-\left(T-T{\s 2}\right)/\tau_{\s s}\big]$, respectively. As a result, the contribution to ${\overline S}_E(T)$ from this realization of the random fields reads

\begin{equation}
\label{echo2}
\int\limits_{0}^{\frac{T}{2}} \frac{dT_{\s 1} e^{-\frac{T_{\s 1}}{\tau_{\s s}}} }{\tau_{\s s}}
\int\limits_{\frac{T}{2}}^{T} \frac{dT_{\s 2}  e^{-\frac{T-T_{\s 2}}{\tau_{\s s}}}}{\tau_{\s s}}
\Big\langle p_{\s 0}( T_{\s 2}-T_{\s 1},\tau)\Big\rangle e^{-b_{\s 0}^2(T-T_{\s 1}-T_{\s 2})^2}.
\end{equation}
To analyze this expression, it is convenient to introduce a new variable, $T_{\s 3}=T-T_{\s 2}$. Then it takes the form
\begin{equation}
\label{echo3}
\int\limits_{0}^{\frac{T}{2}} \frac{dT_{\s 1} e^{-\frac{T_{\s 1}}{\tau_{\s s}}} }{\tau_{\s s}}
\int\limits_{0}^{\frac{T}{2}} \frac{dT_{\s 3}  e^{-\frac{T_{\s 3}}{\tau_{\s s}}}}{\tau_{\s s}}
\Big\langle p_{\s 0}( T-T_{\s 3}-T_{\s 1},\tau)\Big\rangle e^{-b_{\s 0}^2(T_{\s 3}-T_{\s 1})^2}.
\end{equation}
The average in Eq. (\ref{echo3}) is equal to$\newline$ $C_{\alpha}\Gamma(\alpha)\tau_{\s 0}^{\alpha}/\big(T-T_{\s 3}-T_{\s 1}\big)^\alpha$, see Eq. (\ref{28}).

The most sound consequence of Eq. (\ref{echo3}) is that the echo signal survives at times much longer than
$\tau_{\s s}$. Indeed, characteristics $T_{\s 1}$, $T_{\s 3}$ in Eq. (\ref{echo3}) are $\sim \tau_{\s s}$.
For $T\gg\tau_{\s s}$ the upper limits in the integrals can be extended to infinity, while the
average can be replaced by $C_{\alpha}\Gamma(\alpha)\tau_{\s 0}^{\alpha}/T^\alpha$. Then we get
\begin{equation}
\label{echo4}
C_\alpha\Gamma(\alpha)\Big(\frac{\tau_{\s 0}}{T}\Big)^\alpha
\int\limits_{0}^{\infty}dx~ e^{-x}\int\limits_{0}^{\infty}dy~ e^{-y}
\exp\Big[-b_{\s 0}^2\tau_{\s s}^2(x-y)^2\Big],
\end{equation}
where we have introduces the dimensionless variables $x=T_{\s 1}/\tau_{\s s}$ and $y=T_{\s 3}/\tau_{\s s}$.
Within a numerical constant the double integral is equal to $\big(b_{\s 0}\tau_{\s s}\big)^{-1}$. Thus, we
conclude that the echo signal falls off as a power law:

\begin{equation}
\label{echo5}
{\overline S}_E(T)\sim \big(b_{\s 0}\tau_{\s s}\big)^{-1}\Big(\frac{\tau_{\s 0}}{T}\Big)^\alpha.
\end{equation}
It is seen from Eq. (\ref{echo5}) that ${\overline S}_E(T)$ contains a small parameter $\big(b_{\s 0}\tau_{\s s}\big)^{-1}$. At this point, we note that  due to a long tail in the waiting time distribution, there
is a non-exponential probability that during time, $T$, the  carrier does not hop at all.
The contribution to echo signal from such realizations does not contain random magnetic field
and falls with $T$ in the same way as Eq. (\ref{echo5}).
Thus, the final result for the echo signal reads

\begin{equation}
\label{echo6}
{\overline S}_E(T) =\Big\langle p_{\s 0}( T,\tau)\Big\rangle =C_{\alpha}\Gamma(\alpha)\Big(\frac{\tau_{\s 0}}{T}\Big)^\alpha.
\end{equation}

\section{Concluding remarks}

1. As a quantitative measure of acceleration of the relaxation rate, caused by the tail in
the distribution, $F(\tau)$, one can consider a ratio of the times for the values of the tail-parameter $\alpha=1$ and $\alpha=2$. From Eqs. (\ref{19}) and (\ref{25}) one finds
\begin{equation}
\label{R1}
\frac{\tau_{\s s}(\alpha=2)}{\tau_{\s s}(\alpha=1)}=\frac{\pi^{1/2}}{2b_{\s 0}\tau_{\s 0}|\ln(b_{\s 0}\tau_{\s 0})|^2}.
\end{equation}
Both values of $\tau_{\s s}$ are determined by the tail. Parametrical, in $b_{\s 0}\tau_{\s 0}\ll 1$, difference between the two values is
due to the fact that for $\alpha =1$ the portion of sites on which the carrier spin exercises a full rotation
is $\sim (b_{\s 0}\tau_{\s 0})$ for $\alpha =1$ and $\sim (b_{\s 0}\tau_{\s 0})^2$ for $\alpha=2$.

2. In replacing the expression Eq. (\ref{6}) by $\exp(-t/\tau_{\s s})$, we argued that this replacement
is valid for $t\gtrsim b_{\s 0}^{-1}$. This means that in Eq. (\ref{6}) we chose the lower limit $t_1\sim b_{\s 0}^{-1}$.
Uncertainty in this lower limit leads to uncertainty in the prefactor in Eq. (\ref{8}) $ \sim \exp\Big(1/b_{\s 0}\tau_{\s s}\Big)$, which can be neglected since the product $b_{\s 0}\tau_{\s s}$ is big.
Yet another contribution to the prefactor comes from $\int\limits_{\tau_{\s 0}}^{1/b_{\s 0}}dt_1\Phi(t_{\s 1})$.
To estimate this contribution, we note that for $t\lesssim b_{\s 0}^{-1}$
\begin{equation}
\label{R2}
\Phi(t)\Big\vert_{t<b_{\s 0}^{-1}}\approx\frac{1-\int\limits_0^t dt_2\Big\langle\frac{1}{\tau} e^{-t_2/\tau}\Big\rangle}{\int\limits_0^t dt_2t_2\Big\langle\frac{1}{\tau} e^{-t_2/\tau}\Big\rangle}=
\frac{\Big\langle e^{-t/\tau}\Big\rangle}{\Big\langle \tau -(t+\tau)e^{-t/\tau}\Big\rangle}.
\end{equation}
For $t\gtrsim \tau_{\s 0}$ we can neglect the second term in the denominator and perform  integration
over time, yielding

\begin{equation}
\label{R3}
\int\limits_0^t dt_1 \Phi(t_1)\Big\vert_{t<b_{\s 0}^{-1}}=
\frac{\Big\langle\tau(1- e^{-t/\tau})\Big\rangle}{\langle \tau \rangle}.
\end{equation}
Thus, the contribution to the prefactor from small times does not exceed $1$.
In fact, the contribution Eq. (\ref{R3}) comes from  neglecting the first term,
$\big\langle \exp(-\frac{t}{\tau}-b_{\s 0}^2t^2)\big\rangle$,
in the basic equation Eq. (\ref {1}). We effectively replaced the first by the initial condition:
$\overline{S}_z(t\sim\tau_{\s 0})=1$.
More accurate calculation, based on the Laplace transform,
suggests that for $\alpha >1$ the true prefactor is $1$.

3. Formal solution of Eq. (\ref{1}) can be expressed in the form of the inverse Laplace transform,
see e.g. Ref. \onlinecite{Laplace}
\begin{equation}
\label{R4}
{\overline S}_{ z}(t)=\frac{1}{2\pi i}\int\limits_{\gamma-i\infty}^{\gamma+i\infty}
\frac{S_{ z}^{\s (0)}(s)}{1-K(s)}
\end{equation}
where the functions $S_{z}^{\s (0)}(s)$ and $K(s)$ are defined as
\begin{align}
\label{R5}
S_{ z}^{\s (0)}(s)&=\int\limits_0^\infty dt \Big\langle e^{-t/\tau}\Big\rangle e^{-st-b_{\s 0}^2t^2},\\ \nonumber
K(s)&=\int\limits_0^\infty dt \Big\langle\frac{1}{\tau}e^{-t/\tau}\Big\rangle e^{-st-b_{\s 0}^2t^2}
\end{align}
The decay of ${\overline S}_{z}(t)$ is defined by the poles, $s=s_{\s 0}$, for which $K(s_{\s 0})=1$.
For $\alpha >1$, one can retain only the smallest $s_{\s 0}$ and find it by expanding $K(s)$
at small $s$. This readily yields $s_{\s 0}=-\Phi(\infty)=-\tau_{\s s}^{-1}$, i.e. the same expression
Eq. (\ref{9}) for the decay rate as was found in Sect. II from the different approach. The justification
for expanding $K(s)$ is   that the exponent $\exp \big(-b_{\s 0}^2t^2\big)$ ensures the convergence
of the integral Eq. (\ref{R5}) at $t\sim b_{\s 0}^{-1}$ when $\exp\big(-s_{\s 0}t\big)$ is close to $1$.
Thus, for $\alpha >1$, the results of the two approaches to solving Eq. (\ref{1}) coincide. Moreover,
the solution Eq. (\ref{R4}) takes into account the first term in Eq. (\ref{1}), which we have neglected.
The prefactor in front of $\exp \big(-t/\tau_{\s s}\big)$ calculated from Eq. (\ref{R4}) is given by
\begin{equation}
\label{R6}
-\frac{S_{ z}^{\s (0)}(s_{\s 0})}{K'(s_{\s 0})}\approx
\frac{\int\limits_0^\infty dt \Big\langle e^{-t/\tau}\Big\rangle e^{-b_{\s 0}^2t^2}}
{\int\limits_0^\infty dt ~t\Big\langle\frac{1}{\tau}e^{-t/\tau}\Big\rangle e^{-b_{\s 0}^2t^2}}.
\end{equation}
It appears that we can neglect the exponent $\exp\big(-b_{\s 0}^2t^2\big)$ in the integrands in the numerator and the denominator. This is because  both integrals   converge for $\alpha>1$ and are equal to $\langle \tau\rangle$, which is finite for $\alpha>1$.
Thus the true prefactor is equal to $1$, as was mentioned above.

4. For $\alpha <1$ the formal solution Eq. (\ref{R4}) becomes useless. This is because the pole $s_{\s 0}$
cannot be found analytically, and, moreover, many poles (corresponding to $s\sim b_{\s 0}$)  contribute to $\overline{S}_z(t)$. This also follows
from our solution Eq. (\ref{30}) and from Fig. \ref{figure3}. It is seen that
$\overline{S}_z(t)$ follows a simple exponential behavior only for large times, $b_{\s 0}t\gtrsim 1$.

5. In the paper Ref. \onlinecite{Harmon2} the effect  of the power-law tail in $F(\tau)$ on the decay of $\overline{S}_z(t)$ was analyzed. The analysis relied on the solution of Eq.~(\ref{1}) in the form of Eq. (\ref{R4}). The authors
did not analyze the behavior of $\tau_{\s s}$ in different domains of the tail-parameter, $\alpha$.
They rather realized that retaining a singles pole becomes inadequate for $\alpha <1$, and resorted to the numerics.
Our  results  Eqs. (\ref{11}), (\ref{15})   and Eqs. (\ref{19}), (\ref{24}) for the crossover domains are fully analytical. Obtaining these results was facilitated by exploiting the small parameter $b_{\s 0}\tau_{\s 0}\ll 1$ for $\alpha >1$ and by solving Eq. (\ref{1}) using an alternative approach for $\alpha <1$.

\begin{figure}
\includegraphics[width=85mm]{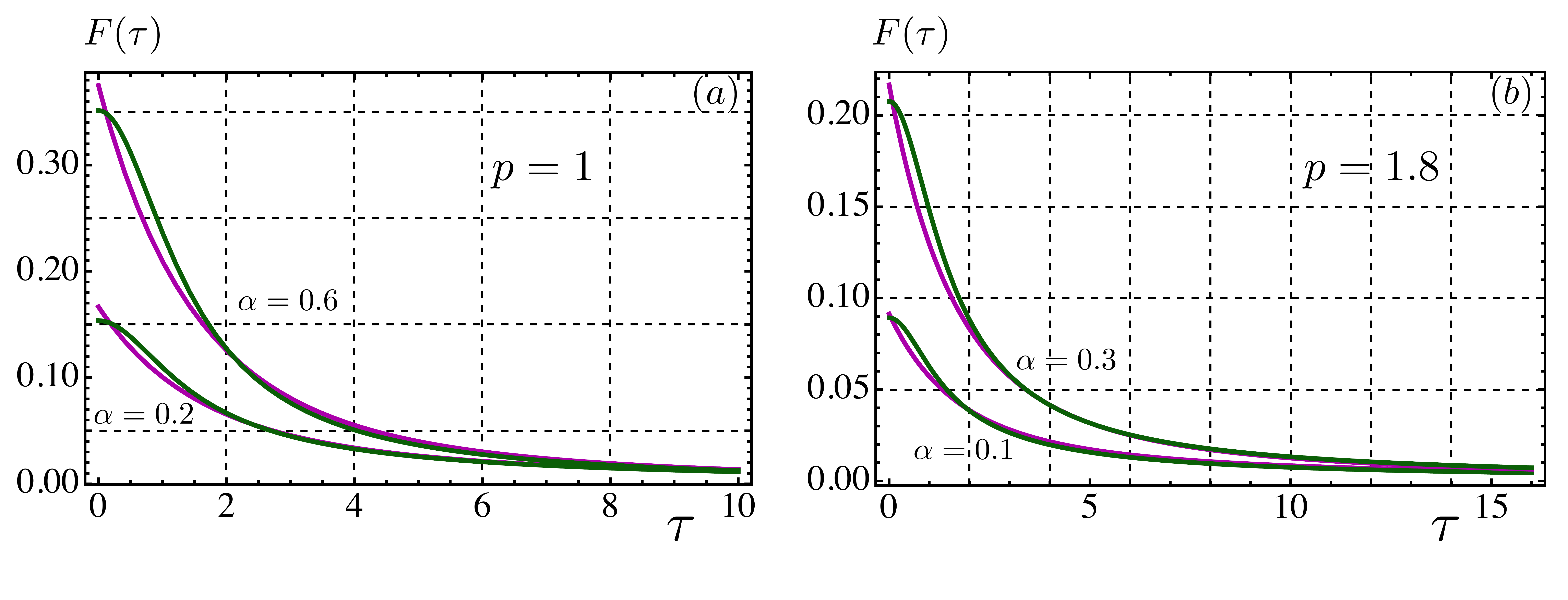}
\caption{(Color online)
Distribution function, $F(\tau)$,  of the waiting times in the multiple-trapping model is shown with purple lines for the densities of tail states of the form   $g(\varepsilon)\propto \exp\left[-\left(|\varepsilon|/{\cal T}_0\right)^p\right]$ with $p=1$ (a) and $p=1.8$ (b). Green lines are the interpolations of $F(\tau)$ with the form Eq. (\ref{F}) of the main text.}
\label{figure5}
\end{figure}

\section{Acknowledgements}

We gratefully acknowledge numerous discussions  with  V. V. Dobrovitski.
 The  work at the University of Utah was supported by NSF MRSEC program
under Grant No. DMR 1121252.
The work at Ames Laboratory was supported by the
Department of Energy – Basic EnergySciences under
Contract No. DE-AC02-07CH11358.

\appendix
\section{Applicability of the waiting times distribution Eq. (\ref{F}) to the multiple trapping model}

In the multiple trapping model\cite{DT1,DT2,DT3,DT4,Harmon1,Harmon2,Baranovskii}, the waiting time is determined by activation of electron from
a localized state in the tail to the conduction band. If the energy position of the localized state is
$-\varepsilon$, then the activation rate is equal to $\Gamma(\varepsilon)=\Gamma_0\exp[\varepsilon/{\cal T}]$, where
${\cal T}$ is the temperature. Actual waiting times, $t_i$, are random. While the average waiting time is $\Gamma^{-1}$, the distribution of the waiting times
for a given $\varepsilon$ is given by the Poisson distribution
\begin{equation}
f_{\varepsilon}(\tau)=\Gamma(\varepsilon)\exp\left[-\Gamma(\varepsilon)\tau\right].
\end{equation}
The remaining task is to average $f_{\varepsilon}(\tau)$ over $\varepsilon$ with the weight determined by the density
of the tail states, $g(\varepsilon)$. In the multiple trapping model the form of $g(\varepsilon)$ is a simple exponent
 $g(\varepsilon)\propto \exp\left[\varepsilon/{\cal T}_0\right]$. The final expression for the waiting times distribution
reads
\begin{equation}
F(\tau)\propto \int\limits_{-\infty}^0 d\varepsilon g(\varepsilon)f_{\varepsilon}(\tau)\propto
\frac{\alpha}{\tau^{\alpha+1}}\int\limits_0^{\Gamma_0\tau}dx ~x^\alpha e^{-x},
\end{equation}
where $\alpha={\cal T}/{\cal T}_0$. For large waiting times $\tau \gg \Gamma_0^{-1}$, we have $F(\tau)\propto \tau^{-(\alpha+1)}$. At $\tau \rightarrow 0$ the power-law divergence is cut off. The character of cutoff is not precisely
the one given by Eq. (\ref{F}), but they match very closely, as illustrated in the Figure \ref{figure5}. In organic semiconductors the density of the tail states is better approximated by a stretched-exponential form $g(\varepsilon)\propto \exp\left[-\left(|\varepsilon|/{\cal T}_0\right)^p\right]$, with $p$ close to $2$, see Refs. \onlinecite{Baranovskii1,Baranovskii2,Robert}. Repeating the above steps for this $g(\varepsilon)$ we
found that $F(\tau)$ can still be closely approximated with Eq. (\ref{F}), see Figure \ref{figure5}.

%interpolation of $F(\tau)$ with Eq. [5] of the main text is

\end{document}